\documentclass[a4paper]{article}
\usepackage{graphicx}
\usepackage{amsmath,amssymb}
\usepackage[title]{appendix}
\begin{document}
\large
\title{The evolution of localized vortex in stably stratified flows}
\author{Levinski Vladimir}
\maketitle
\pagenumbering{arabic}
\section{Abstract}

The evolution of a localized vortex in stably stratified flow, within the Boussinesq approximation, is analyzed using the fluid impulse concept. The set of equations describing the temporal development of the fluid impulse has an integro-differential character where the terms representing the effect of stratification appear as convolution integral of the component of the fluid impulse and time-depended 'memory' functions. These functions are calculated for the case where the external parallel shear flow
varies only in the direction gravitational force and is subjected to localized two- and three-dimensional disturbances. As follows from the solution of evolution equations, in both cases there is a range of Richardson numbers where the fluid impulse associated with the disturbance grows exponentially. The upper limit of this range for two- and three-dimensional cases are {\itshape Ri}$\; \sim \;$1.23 and {\itshape Ri}$\;\sim \;$0.89. Both cases are also characterized by a critical value of the Richardson number (around {\itshape Ri}$\;\sim \;$0.3 for both cases), beyond which the solution exhibits oscillatory behavior. Indeed, this oscillatory behavior has been observed in turbulent flows and, as is shown in the present study, it is an inherent feature of a {\itshape non-wavy localized vortex} embedded in a stably stratified shear flow. The paper was written in 2001 and published now without changes and new additions.

\section{Introduction}

Because of their direct applications on geophysical flows, turbulent shear flows under the influence of a stable stratification have been an attractive subject of research from both scientific and engineering points of view. Qualitatively, the behavior of stably stratified shear flows can be classified into two typical categories: shear and buoyancy dominated flows (see the brief review in Holt, Koseff $\And$ Ferziger 1992). The relative importance of these two effects can be measured by the non-dimensional Richardson number ({\itshape Ri}). In general, when  {\itshape Ri}$\; > \;$0.25, the flow is dominated by stratification, whereas for  {\itshape Ri}$\; < \;$0.25, the flow is shear dominated.

In their experiments of stably stratified shear flow, Rohr {\itshape et al.} (1988) observed that the Richardson number at which turbulence growth was suppressed is {\itshape Ri}$\; = \;$0.25$\pm $0.05. The same value was obtained by Holt {\itshape et al.} (1992) in their numerical simulations, provided that the Reynolds number was efficiently large. Moreover, for {\itshape Ri}$\; > \;$0.25, Holt {\itshape et al.} (1992) found that the average vertical density flux approaches zero. The value of {\itshape Ri}$\; = \;$0.25 is in fact the critical value for stability predicted by the inviscid linear analysis (Miles 1961). However, as was discussed by  Holt {\itshape et al.} (1992) and Rohr {\itshape et al.} (1988), the relevance of the results obtained by linear stability analysis for turbulent shear flows where significant nonlinear disturbances are present is questionable.

The experimental detection of well organized localized vortex structures in turbulent boundary layers (Kline {\itshape et al.} 1967, Head $\And$ Bandyopadhyay 1981) has made it possible to gain insight into the mechanisms that govern these flows. Using data bases obtained by direct numerical simulations, similar structures, having the shape of a hairpin, were found in fully developed turbulent channel flow (Moin $\And$ Kim 1985) and homogeneous turbulent shear flow (Roges $\And$ Moin 1987). In the numerical simulations of stably stratified shear flows by Gertz (1991a,b) and Holt {\itshape et al.} (1992), hairpin vortices were clearly identified at subcritical Richardson numbers (shear-dominated flows). For weakly supercritical buoyancy-dominated flows, Gertz (1991a) observed the transformation of hairpin vortices into vortex rings with various orientations. These localized vortex structures completely disappeared when the Richardson number was increased above 1 (Gertz 1991b, Holt {\itshape et al.} 1992). The above mentioned features suggest that the effect of a stable stratification on the nature turbulent shear flows is strongly correlated with the effect on the vortical structure observed in these flows. This correlation is the main motivation behind the present study.

The objective of the present work was to investigate the effect of a stable stratification on the evolution of the finite amplitude localized disturbance in shear flow. It is based on the theoretical approach described in Levinski $\And$ Cohen (1995, hereinafter referred to LC). They considered the time evolution of a localized vortex disturbance, all dimensions of which are much smaller than the length scale characterizing variations of the basic (unperturbed) fluid-velocity gradient, and used the fluid impulse to characterize such a disturbance. The fluid impulse definition is given by

\begin{equation}
\mathbf{I}=\frac{1}{2} \int \mathbf{x} \times \boldsymbol{\omega}(\mathbf{x}) dV , 
\end{equation}
where $\mathbf{x}$ is the position vector, $\boldsymbol{\omega}$ is the disturbance vorticity vector, $dV$ is a volume element and the integration extends over the entire fluid domain.

In unbounded flows the fluid impulse integral is not modified by self-induced motion (Batchelor, 1967). Consequently, its evolution satisfies a linear equations, even though the fluid motion itself is governed by inherently non-linear effects. It is this property that makes fluid impulse formalism so attractive. While the integral character  of the fluid impulse does not provide the details of flow within the disturbed vortical region, this insensitivity yields, in turn, some universal properties.

Turner (1967, 1960) was the first to use the fluid impulse integral for describing the effect of buoyancy on the evolution of the vortex rings and vortex pairs. However, as was pointed out by Saffman (1972), the new vorticity field generated by the motion of the vortex disturbance was treated in Turner's analysis in a rough way. Score $\And$ Davenport (1970) considered the motion of a vortex pair in a stratified atmosphere using the fluid impulse approach. But still the dynamic effect of the new generated vorticity was neglected. Saffman (1972) considered the application of the fluid impulse to stratified flows. He showed that the concept of the fluid impulse is useful, providing the density differences are small and the Boussinesq approximation can be made. Nevertheless, in his description of vortex pair motion in a stratified atmosphere, it was approximated as a sequence of states where the motion at any instant is like that of vortex pair through a uniform fluid. In the present work, the use of Saffman's approximation is not necessary. Based n the method proposed by LC, it is shown that the evolution of a vortex disturbance can be described in terms of the fluid impulse without any additional assumptions, provided that the Boussinesq approximation holds and the initial disturbance is compact. In section 2, the general formulation of the model is described. The results for three-dimensional disturbance (e.g. hairpin vortex) and two-dimensional disturbance (e.g. vortex pair) are presented in subsections 3.1 and 3.2 respectively.

\section{Analysis}

The flow field is considered as a sum of the basic (unperturbed) field and a finite amplitude disturbance field. Consequently, the total field can be represented as

\begin{equation*}
\mathbf{U}_{T}=\mathbf{U}_{0}(z)+\mathbf{u}(x,y,z,t) \;, \; \rho_{T}=\rho_{0}(z)+\rho(x,y,z,t) \;, \; P_{T}=P_{0}(z)+p(x,y,z,t)
\end{equation*}
where $\mathbf{U}_{T}, \;\mathbf{U}_{0}$ and $\mathbf{u}$ are the respective total, base and perturbed velocity fields, $ \rho_{T},\;\rho_{0}$ and $\rho$ the respective densities and $P_{T},\;P_{0}$ and $p$ the respective pressure fields. We use the Cartesian coordinate system ${x,y,z}$, where the gravity vector $\mathbf{g}$ points in the negative z-direction. The base flow is assumed to be given by

\begin{equation*}
\mathbf{U}_{0}=(U_0(z),0,0), \; \rho_{0}=\rho_{0}(z) \;  and \; P_{0}(z)=P_{0}(0)- g\int_0^z \rho_0(z')dz'
\end{equation*}

In most situations, the scale of density variations is substantially larger than that of velocity. In this case, the flow field can be approximated by the incompressible condition. Furthermore, the density perturbation can be assumed as small even though the velocity perturbation has a finite amplitude. Then, using the Boussinesq approximation, where the effect of density variation is retained only in buoyancy force, the equations of motion, continuity and incompressibility can be written as

\begin{equation}
\frac{\partial \mathbf{U}_{T}}{\partial t}+ (\mathbf{U}_{T} \cdot \mathbf{\nabla}) \mathbf{U}_{T}=-\frac{\mathbf{\nabla} P_{T}}{\rho_0}+\mathbf{g} \frac{\rho_T-\rho_0}{\rho_0}
\end{equation}
\begin{equation}
\frac{\partial \rho_{T}}{\partial t}+ (\mathbf{U}_{T} \cdot \mathbf{\nabla}) \rho_{T}=0
\end{equation}
\begin{equation}
\mathbf{\nabla} \cdot \mathbf{U}_{T}=0
\end{equation}

To complete the problem formulation, Eqs (2)-(4) need to be supplemented with the appropriate initial conditions for the disturbed density and vorticity fields. We consider the case of the flow subjected to a local vortex disturbance assuming the initial condition for disturbed density to be zero. The initial vortex disturbance is not required to have an infinitesimally small amplitude but it is assumed to be confined within a small region with typical scale much smaller than the length scale characterizing the base velocity field.

It should be emphasized that the formulation given above contains two distinctions from the conventional statement of the problem for stability analysis. The first one is the requirement of compactness of the vortex, e.g. the initial disturbance is assumed to have sufficiently non-wavy character. This approach accounts for the nature of hairpin vortices observed in boundary layers as well as in stably stratified shear flows for  {\itshape Ri} $ \leq \;$1. In this paper we are concerned with analysis of conditions for localized disturbance initiation. This initiation can be performed in a variety of ways including specific to the stratified flows method (Andreassen {\itshape et al.}, 1998), where creating hairpin vortices was caused by internal gravity wave breaking.

The second distinction is that we do not assume the amplitude of the disturbance to be small. This is of fundamental importance because experimental results concerning stability of shear flows with respect to localized vortex disturbances show (Masihito $\And$ Nishioka, 1995, Malkiel, Levinski $\And$ Cohen, 1999) that for instability to occur a sufficient level of initial disturbance amplitude is required.

\subsection{Three-dimensional case}
\subsubsection{General consideration}

We characterize the vortex disturbance by the fluid impulse integral $\mathbf{I}$ defined in (1). The corresponding time derivative is given by

\begin{equation}
\frac{d\mathbf{I}}{dt}=\frac{1}{2} \int \mathbf{x} \times \frac{\partial \boldsymbol{\omega}(\mathbf{x,t})}{\partial t} dV , 
\end{equation}
where the time derivative of the disturbed vorticity is obtained applying operator rotor to (2) and subtracting the undisturbed vorticity equation, i.e.

\begin{equation}
\frac{\partial \boldsymbol{\omega}}{\partial t}+ (\mathbf{U}_{0} \cdot \mathbf{\nabla}) \boldsymbol{\omega}-(\boldsymbol{\omega} \cdot \mathbf{\nabla}) \mathbf{U}_{0}-(\boldsymbol{\Omega}_0 \cdot \mathbf{\nabla}) \mathbf{u}+ (\mathbf{u} \cdot \mathbf{\nabla}) \boldsymbol{\omega}-(\boldsymbol{\omega} \cdot \mathbf{\nabla}) \mathbf{u}=\frac{1}{\rho_0} \mathbf{\nabla} \rho \times \mathbf{g}, 
\end{equation}
where $\boldsymbol{\Omega}_0= \mathbf{\nabla} \times \mathbf{U}_{0}$. In (6) we also make use of the compactness of the disturbance and neglect terms which have second derivative of the basic velocity and pressure fields. This means that the undisturbed basic flow, which is assumed to be known, is approximated by the leading terms of the Taylor series expansion, i.e. $\boldsymbol{\Omega}_0$ is constant and $U_0(z)=U_0(0)+\Omega_0 z$. Note that the fluid impulse is invariant under Galilean transformation, and we can set  $U_0(0)=0$ without loss of generality. Although the leading term of $\rho_0$ in (6) is a constant, its variation with $z$ is the only source for generating the density disturbance. The corresponding evolution equation is 
\begin{equation}
\frac{\partial \rho}{\partial t}+ (\mathbf{U}_{0} \cdot \mathbf{\nabla}) \rho+ (\mathbf{u} \cdot \mathbf{\nabla}) \rho+u_z \frac{d\rho_0}{dz}=0, 
\end{equation}
where $d\rho_0/dz$ is constant to the leading order.

As mentioned above the main advantage of using the fluid impulse is that its evolution is described by a linear equation, even though the fluid motion itself is strongly non-linear. However, the use of fluid impulse is not a straightforward procedure since, as it explained in the following, the integral (on the right hand side of (5)) is not absolutely convergent. The reason for this is the generation of far field vorticity perturbations (tails) caused by the velocity field induced by the localized vortical disturbance. For example, the far field velocity induced by the localized disturbance at a distance $|\mathbf{x}|>> l$ is of order of $|\mathbf{u}|\sim  |\mathbf{I}|/|\mathbf{x}|^3$, where $\mathbf{I}$ is the magnitude of the associated fluid impulse integral and $l$ represents the typical length scale of the disturbance. From (7) it follows that the disturbed density is of the order of $|\rho| \sim \mathcal{O}(1/|\mathbf{x}|^3)$. Correspondingly, the fourth term on the left hand side and the term on the right hand side of (6) (hereinafter referred to as problematic terms), which are responsible for the generation of the new vorticity over the entire flow region, exhibit asymptotic behavior of $\mathcal{O}(1/|\mathbf{x}|^4)$. Substitution of these terms into (5) causes the integral to be not absolutely convergent, i.e. its evolution depends on how the volume of integration is allowed to tend to infinity.

To overcome this difficulty we follow the approach proposed in LC, which makes it possible to distinguish the contribution to the fluid impulse integral of the 'local' vorticity field from that of the far-field tails, Accordingly, we represent the velocity, vorticity and density fields as a superposition of two components $\boldsymbol{\omega}= \boldsymbol{\omega}^I+ \boldsymbol{\omega}^{II}$, $\mathbf{u}=\mathbf{u}^I+\mathbf{u}^{II}$ and $\rho=\rho^I+\rho^{II}$, so that $\boldsymbol{\omega}^{I,II}=\mathbf{\nabla} \times \mathbf{u}^{I,II}$. For each component we require that

\begin{equation}
\mathbf{\nabla} \cdot \boldsymbol{\omega}^I=\mathbf{\nabla} \cdot \boldsymbol{\omega}^{II} =0
\end{equation}
The fist component, indicated by the superscript $I$, is associated with the concentrated vorticity and density distributions confined within and in the vicinity of the initially disturbed region, whereas the second component, indicated by the superscript $II$, is associated with the far-field vorticity and density generated by the problematic terms in (6) and (7). Consequently, we set the initial distribution of the vorticity and density fields as

\begin{equation*}
\boldsymbol{\omega}^I(\mathbf{x},t=0)= \boldsymbol{\omega}(\mathbf{x},t=0), \; \boldsymbol{\omega}^{II}(\mathbf{x},t=0)=0 \; and
\end{equation*}
\begin{equation}
 \rho^I(\mathbf{x},t=0)=\rho^{II}(\mathbf{x},t=0)=0
\end{equation}
and we intend to follow the evolution of the fluid impulse associated with of $\boldsymbol{\omega}^I(\mathbf{x},t)$ alone. Accordingly we redefine the fluid impulse of the localized disturbance as 

\begin{equation}
\mathbf{I}=\frac{1}{2} \int \mathbf{x} \times \boldsymbol{\omega}^I(\mathbf{x},t) dV 
\end{equation}

As is shown in Appendix A, the subdivision procedure has resulted in an evolution equation for $\boldsymbol{\omega}^I$ vorticity field with asymptotic behavior 

\begin{equation}
|\boldsymbol{\omega}^I(\mathbf{x},t)| \sim \mathcal{O}(\frac{1}{|\mathbf{x}|^5})  \;\; for \; |\mathbf{x}|>> l
\end{equation}

Before proceeding with the analysis we would like to note that the tails of the vorticity field $\boldsymbol{\omega}^I$ can be completely eliminated by canceling the subsequent members in far field asymptotic expansion of the problematic terms. In this case, the corresponding vorticity field $\boldsymbol{\omega}^{II}$ will include total far field vorticity tails and a vorticity field within the sphere to close the vortex lines of the $\boldsymbol{\omega}^{II}$ field which intersect the sphere surface from the outer region (see Figure 1 below)

\begin{figure}[h]
\centering
\includegraphics[scale=0.6]{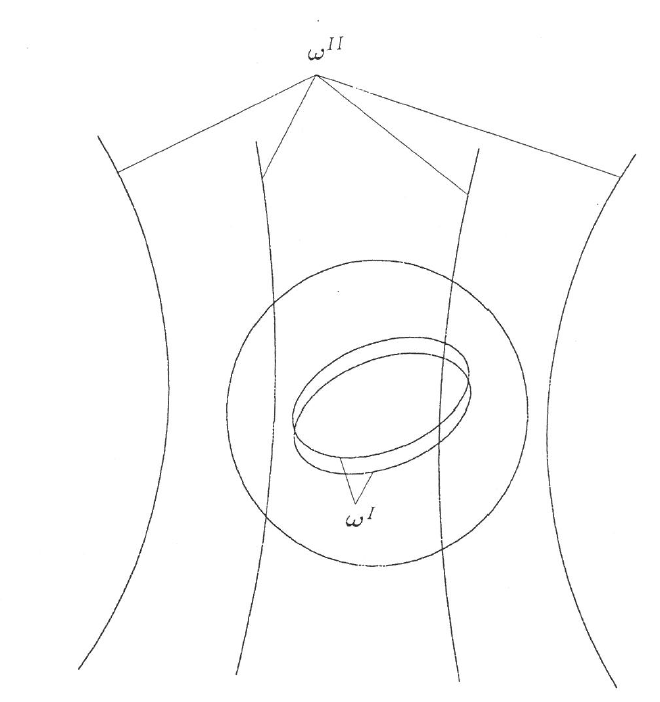}
\end{figure}

Figure 1. Schematic view of the  $\boldsymbol{\omega}^I$ and $\boldsymbol{\omega}^{II}$ vorticity fields. The sum of $\boldsymbol{\omega}^I + \boldsymbol{\omega}^{II}$ corresponds to the total vorticity field of the disturbance.
\\
\\
It is also follows from the subdivision procedure that the initially zero vorticity field $\boldsymbol{\omega}^{II}$ and the associated density field $\rho^{II}$ are generated within the sphere only due to the artificial term $\mathbf{\nabla} \Psi$ in the right hand side of equation (A8). Since $\Psi$ and their derivatives are harmonic functions, $\mathbf{\nabla} \Psi$ attains its maximum on the boundary of the sphere, where its magnitude can be estimated from (A11) as $\Omega_0 |\mathbf{I}|/R^4$. Consequently, the order of magnitude of the vorticity $\boldsymbol{\omega}^{II}$ generated within the sphere is $|\mathbf{I}|/R^4$. On the other hand, the localized vortex of hairpin type observed in boundary layer flows possesses a very concentrated vorticity field. It is confined within vortex tubes with the typical cross-section diameter $l_1 << l$. Correspondingly, the magnitude of fluid impulse  $|\mathbf{I}|$ associated with $\boldsymbol{\omega}^I$ field can be estimated as $|\mathbf{I}|=\mathcal{O}(|\boldsymbol{\omega}^I|l^2l_1^2)$. Thus the ratio $|\boldsymbol{\omega}^{II}|/|\boldsymbol{\omega}^I|$ within the inner region $|\mathbf{x}| \leq R$ is proportional to the small value $l_1^2/l^2$.

According to (11) the fluid impulse integral (10) is absolutely convergent. It is therefore possible to use an infinite sphere as the volume of integration. Consequently, the time evolution of $\mathbf{I}$ is given by

\begin{equation}
\frac{d\mathbf{I}}{dt}=\frac{1}{2} \lim_{R_1 \to \infty} \int_{|\mathbf{x}| \leq R_1} \mathbf{x} \times \frac{\partial \boldsymbol{\omega}(\mathbf{x,t})}{\partial t} dV  
\end{equation}

Straightforward calculations show that the contribution to the integral on the right hand side of (12) from the terms containing $\rho^a$ and $\mathbf{u}^a$ in (A1) as well as from the term $\mathbf{\nabla} \Psi$  in (A7) is zero. Thus, these artificial terms, used to extract the localized part of the disturbance, have no direct impact on the evolution of the fluid impulse associated with the concentrated vorticity  $\boldsymbol{\omega}^{I}$, provided that the volume of integration in fluid impulse definition (12) is an infinite sphere. The resulting evolution equation for fluid impulse is

\begin{equation*}
\frac{d\mathbf{I}}{dt}=-\frac{1}{2} \lim_{R_1 \to \infty} \int_{|\mathbf{x}| \leq R_1} \mathbf{x} \times [ (\mathbf{U}_{0} \cdot \mathbf{\nabla}) \boldsymbol{\omega}^I-(\boldsymbol{\omega}^I \cdot \mathbf{\nabla}) \mathbf{U}_{0}-(\boldsymbol{\Omega}_0 \cdot \mathbf{\nabla}) \mathbf{u}^I+ (\mathbf{u} \cdot \mathbf{\nabla}) \boldsymbol{\omega}-(\boldsymbol{\omega} \cdot \mathbf{\nabla}) \mathbf{u}] dV  
\end{equation*}
\begin{equation}
+\frac{1}{2\rho_0} \lim_{R_1 \to \infty} \int_{|\mathbf{x}| \leq R_1} \mathbf{x} \times [\mathbf{\nabla} \rho^I \times \mathbf{g}]dV  
\end{equation}

\subsubsection{Solution procedure}

The first integral in (13) is identical to the one calculated by LC for the case when the basic flow has a constant density. The calculation of the second integral is given in Apendix B. Finally, the evolution of the fluid impulse components is described by the following set of linear equations

\begin{equation}
\frac{dI_x}{dt}=-\frac{1}{2}I_z+Ri \int_0^{\bar{t}}[G_{xx}(\bar{t}-\bar{\tau})I_x(\bar{\tau})+G_{xz}(\bar{t}-\bar{\tau})I_z(\bar{\tau})] d\bar{\tau}
\end{equation}
\begin{equation}
\frac{dI_y}{dt}=Ri \int_0^{\bar{t}}G_{yy}(\bar{t}-\bar{\tau})I_y(\bar{\tau}) d\bar{\tau}
\end{equation}
\begin{equation*}
\frac{dI_z}{dt}=-\frac{1}{2}I_x-Ri \int_0^{\bar{t}}[G_{zx}(\bar{t}-\bar{\tau})I_x(\bar{\tau})+G_{zz}(\bar{t}-\bar{\tau})I_z(\bar{\tau})] d\bar{\tau}-\frac{2}{3} Ri  \int_0^{\bar{t}} I_z(\bar{\tau}) d\bar{\tau}-
\end{equation*}
\begin{equation}
2Ri \int_0^{\bar{t}} d\bar{\tau} \int_0^{\bar{\tau}}[G_{xx}(\bar{\tau}-\bar{\nu})I_x(\bar{\nu})+G_{xz}(\bar{\tau}-\bar{\nu})I_z(\bar{\nu})] d\bar{\nu}
\end{equation}
where the 'memory' functions $G_{xx}, G_{yy}, G_{xz}, G_{zx} \;and \; G_{zz}$ are calculated in Apendix B (and shown below in Figure 2). Here $\bar{t}=\Omega_0 t$ is the nondimensional time, $Ri=\frac{N^2}{\Omega_0^2}$ is the Richardson number and $N$ is the Brunt-V$\ddot{a}$is$\ddot{a}$l$\ddot{a}$ frequency defined as $N^2=-\frac{g}{\rho_0}\frac{d\rho_0}{dz}$.

\begin{figure}[h]
\centering
\includegraphics[scale=0.95]{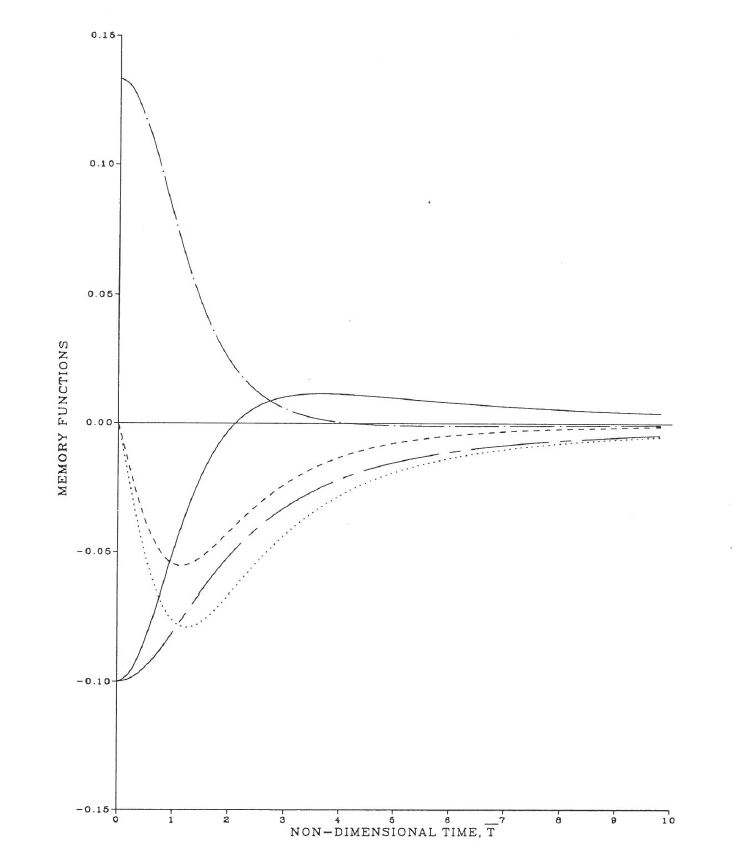}
\end{figure}

Figure 2. The dependence of the 'memory' functions on the non-dimensional time $\bar{t}=\Omega_0 t$. $G_{xx}$, solid line;  $G_{xz}$, dotted line;  $G_{zx}$, dashed line;  $G_{zz}$, chain-dotted line;  $G_{yy}$, chain-dashed line.
\\
\\
Because $G_{yy}$ is negative for all $\bar{t}$, (15) contains only a not-growing solution. In addition, $I_y$ does not depend on $I_x$ and $I_z$ and vice-versa, and therefore, without loss of generality, $I_y$ could be set to zero.

The system (14), (16), is solved using Laplas transform with espect to $\bar{t}$. Accordingly, we define

\[
\binom{I_{x,z}(s)}{G_{ij}(s)}=\int_0^{\infty} d\bar{t} \binom{I_{x,z}(\bar{t})}{G_{ij}(\bar{t}} e^{-s\bar{t}}
\]
where for values of $Re(s)<0$ the functions ${G_{ij}(s)}$ are calculated by extending the corresponding solutions obtained for 
$Re(s) \geq 0$ using analytical continuation. Thus, the solutions of the tansformed equations are 

\begin{equation}
I_x(s)=[A_{zx}I_x(0)+A_{xx}I_z(0)]/D, \;\; I_z(s)=[A_{zz}I_x(0)+A_{xz}I_z(0)]/D,
\end{equation}
where $I_x(0)$ and $I_z(0)$ are the fluid impulse components at time $\bar{t}=0$, $D=A_{xx}A_{zz}-A_{xz}A_{zx}$ and

\begin{equation*}
A_{xx}=s-Ri \cdot G_{xx}(s), \;\; A_{xz}=-\frac{1}{2}+Ri \cdot G_{xz}(s),
\end{equation*}
\begin{equation*}
A_{zx}=-\frac{1}{2}-Ri \cdot G_{zx}(s)-\frac{2Ri}{s}G_{xx}(s), 
\end{equation*}
\begin{equation*}
A_{zz}=s+Ri \cdot G_{zz}(s)+\frac{2}{3}\frac{Ri}{s}+\frac{2Ri}{s}G_{xz}(s)
\end{equation*}

To obtain the temporal behavior of the components of the fluid impulse, it is sufficient to invert the Laplas transform. The inverse is given by

\begin{equation}
I_{x,z}(\bar{t})=\frac{1}{2\pi i} \int_C e^{s\bar{t}} I_{x,z}(s) ds,
\end{equation}
where the contur $C$ in the complex $s$ plane is located to the right of all singularities of the integrand. The evaluation of the integral in (18) is performed using the residues method. The result obtained here includes only the contributions from zeros of $D$.

\subsubsection{Richardson number classification of the solution}

We focus on the solutions of equation $D(s)=0$ with $Re(s)>0$ which correspond to time growing disturbances. The dependence of such eigenvalues (with a positive real part) on the Richardson number is shown in Figure 3. 

\begin{figure}[h]
\centering
\includegraphics[scale=1.2]{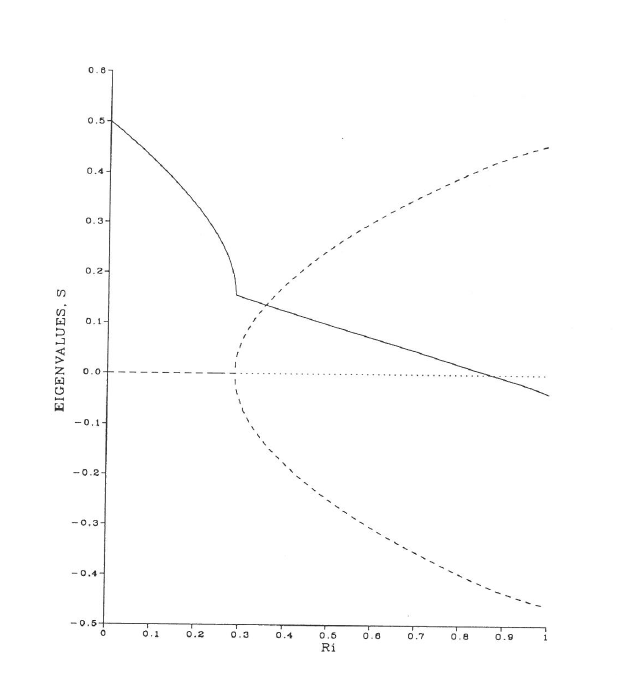}
\end{figure}

Figure 3. The dependence of the eigenvalues, which correspond to time growing disturbances, on the Richardson number for the three-dimensional case. The real part of the eigenvalue ($Re(s)$) is shown by the solid curve while the dash line corresponds to its imaginary part ($Im(s)$). The dotted line corresponds to the marginal value for stability.
\\
\\
For values of $Ri < 0.29$ there is only single growing (in time) solution with a zero imaginary part. For $0.29<Ri<0.88$ two conjugated growing solutions exist. For $Ri>0.88$ the flow is stable with respect to localized disturbance. This behavior is typical for buoyancy dominated flows and corresponds to the conversion of the vertical kinetic energy int the potential energy in the gravitational field.

The evolution equations (14)-(16)  for fluid impulse components are unsuitable for analysis of the vortex disturbance behaviour at $Ri \to \infty$. Formally, this is due to using $1/|\boldsymbol{\Omega}|$ as a relevant time scale to render equations (14)-(16) dimensionless, whereas in the case of large Richardson numbers the characteristic time scale is associated with the effect of the buoyancy forces. However, the main reason that the analysis of (14)-(16) is limited to relatively small Richardson numbers lies in the assumption that the vorticity field $\boldsymbol{\omega}^I$ associated with localized disturbance is strongly concentrated. According to the model, the disturbance is considered to be a compact vortex ($|\boldsymbol{\omega}^{II}|<<|\boldsymbol{\omega}^I|$ within the disturbed region). Here the strength of the vorticity tails rapidly decreases with distance fom the disturbance.

As was shown by LC, the mechanism of hairpin vortex formation includes generation of the new vorticity in the desturbed region and its surrounding. Because the rate of generation as its maximum in the region adjacent to the vortex core, it is bounded to cause an unabated vortex core expansion. The stretching of the new generated vorticity by external shear flow makes it possible to hold a high concentration of vorticity, which is typical for hairpin vortices in turbulent and laminar boundary layers. Gerz (1991a, 1991b) and  Holt {\itshape et al.} (1992) demonstrated that the vortices of this type are formed in stably stratified shear flows in regions of Richardson numbers $Ri \leq 1$. By this means that the application area of the given approach corresponds to the region of Richardson numbers for which the flow is found to be unstable with respect to localized vortex disturbances.

The region of instability can be divided into two domains with different character. For values of $Ri < 0.29$, the solution describes a growing (in time) hairpin type vortex. For example, in the case of non-stratified flow ($Ri=0$) we recover the same solution ($s=1/2$) which was obtained by LC, associated with the growth of a vortex inclined at 45$^0$ to the direction of the basic flow. For values of $Ri>0.29$, the solution represent the product of the term describing the exponential growth and the term describing the periodical motion. For symmetrical vortex structures, such as vortex rings, the direction of fluid impulse of the vortex coincides with the direction of its self-induced motion (Batchelor, 1967). For non-symmetrical vortices this is not always the case, however, it can be shown that the direction of self-induced velocity, averaged over the spherical volume containing the vortex, coincides with the direction of its fluid impulse (see Appendix A in LC). Correspondingly, the periodical term in the solution fr fluid impulse describes an oscillatory motion of the whole vortex disturbance about the initial state.

Behaviour of this type has been observed in several experimental and numerical studies for values of $Ri > 0.25$ and was previously associated with the presence of internal waves (Steward 1969, Stillinger {\itshape et al.} 1983). However, this point of view was not supported by subsequent studies (see Rohr {\itshape et al.} 1988 and Holt {\itshape et al.} 1992 for more detailed discussion). The results of the present study makes it possible to explain the oscillatory behavior without invoking the internal wave concept.

Additional quantitative comparison with results reported by Holt {\itshape et al.} (1992) and Kaltenbach, Gerz and Schuman (1994) for region of $Ri > 0.25$ could be made. They carried out two sets of calculations at $Ri=0.5$ and $Ri=1$. In both cases, oscillating behavior of the flow parameters was observed. We will use the results for the case of $Ri=0.5$ because it falls within the region of instability predicted here. Holt {\itshape et al.} (1992) and Kaltenbach {\itshape et al.} (1994) presented the temporal dependence of ensemble averaged vertical density flux $\overline{\rho v}$ (Figure 15 and Figure 3(b) correspondingly), where $v$ is the disturbed vertical velocity, and $\rho$ is the disturbed density. They found that the zero crossing of $\overline{\rho v}$ occurs at non-dimensional times $\Omega t \simeq 5$ and $\Omega t \simeq 4.5$, respectively. On the other hand, according to the model, the time between zero crossing corresponds to a quater of the oscillation period ($\rho$ is equal to zero at the beginning and at the half of the period, when the vortex is in its equilibrium position in the density gradient, whereas $v$ is equal to zero at the quarter and three quarters of the period, when the vortex reverses the direction of its motion due to buoyancy forces). The result, which can be obtained from Figure 3 for the quarter of the non-dimensional period, is $\Omega T/4=\pi/2Im(s) \simeq 6$. Even though we compare the results of turbulent flow numerical simulations with prediction of laminar inviscid model, the agreement is fairly good.

\subsection{Application to two-dimensional case}

\subsubsection{General consideration}

In this section we use the fluid impulse concept to describe the motion of a vortex pair through a stratified environment. This problem has been studied by many investigators, partially because of its relevance to practical needs.The common example is the hazard to following aircraft caused by the trailing vortex system of a large aircraft.

In two-dimensional flows, the fluid impulse per unit length can be written as 

\begin{equation}
\mathbf{I}=\int \mathbf{r} \times \boldsymbol{\omega}(\mathbf{r}) dA , 
\end{equation}
where $\boldsymbol{\omega}=(0, \omega,0), \; \mathbf{r}=(x,z), \;dA=dxdz$. In this case equation (6) describing the time evolution of the disturbed vorticity is reduced to

\begin{equation}
\frac{\partial \omega}{\partial t}+ (\mathbf{U}_{0} \cdot \mathbf{\nabla}) \omega+ (\mathbf{u} \cdot \mathbf{\nabla}) \omega=\frac{g}{\rho_0} \frac{\partial \rho}{\partial x}
\end{equation}
The evolution equation for the disturbed density remains the same as (7).

As can be seen from (20), the only term responsible for the generation of new vorticity is the buoyancy term in the right hand side. That is, in the case where the buoyancy mass moves in homogeneous ambient fluid (the case considered by Saffman, 1972), the integral (19) is well defined for all times. However, when the surrounding fluid is not homogeneous, the asymptotic behavior of the disturbed vorticity ('tails') is $|\omega| \sim \mathcal{O}(1/|\mathbf{r}|^3)$ and consequently the fluid impulse integral (19) is not absolutely convergent.

To solve this problem, which is similar to the one discussed in the previous section, we use the same procedure and subdivide the vorticity and density fields into two components: the localized fields denoted by the subscript $I$ and the far-fields denoted by the subscript $II$. As for the three-dimensional case, the initial conditions are specified by (9) and we use the subdivision of the plane into two regions, inside and outside circle enclosing the disturbance. urther we follow the sequence of operation given in Appendix A, which in this case is more straightforward because of the divergence free requirement (8) for the two-dimensional case is always fulfilled. Namely, there is no need for artificial generation of the vrticity field $\boldsymbol{\omega}^{II}$ within the circle as is in the three-dimensional case.

According to the procedure, we define the fluid impulse associated with the localized part of the disturbance by 

\begin{equation}
\mathbf{I}=\lim_{R \to \infty} \int_{|\mathbf{r}| \leq R} \mathbf{r} \times \boldsymbol{\omega}(\mathbf{r},t) dA 
\end{equation}
Corresponding evolution equation for $\mathbf{I}$ can be represented as

\begin{equation}
\frac{d\mathbf{I}}{dt}=\lim_{R \to \infty} \int_{|\mathbf{r}| \leq R} ( \mathbf{e}_x z- \mathbf{e}_z x) [(\mathbf{U}_{0} \cdot \mathbf{\nabla}) \omega^I-\frac{g}{\rho_0} \frac{\partial \rho^I}{\partial x}] dA ,
\end{equation}
where $\mathbf{e}_x$ and $\mathbf{e}_z$ are the unit vectors in the $x$ and $z$ directions.

Then, in the same fashion as in Appendix B, the following set of equations describing the time evolution of the fluid impulse components can be obtained

\begin{equation}
\frac{dI_x}{d\bar{t}}=-2 Ri \int_0^{\bar{t}}[G_1(\bar{t}-\bar{\tau})I_z(\bar{\tau})+G_2(\bar{t}-\bar{\tau})I_x(\bar{\tau})] d\bar{\tau}
\end{equation}
\begin{equation*}
\frac{dI_z}{d\bar{t}}=-2 I_x-2 Ri \int_0^{\bar{t}}[-G_1(\bar{t}-\bar{\tau})I_x(\bar{\tau})+G_2(\bar{t}-\bar{\tau})I_z(\bar{\tau})] d\bar{\tau}-2 Ri  \int_0^{\bar{t}} I_z(\bar{\tau}) d\bar{\tau}+
\end{equation*}
\begin{equation}
4 Ri \int_0^{\bar{t}} d\bar{\tau} \int_0^{\bar{\tau}}[G_1(\bar{\tau}-\bar{\nu})I_z(\bar{\nu})+G_2(\bar{\tau}-\bar{\nu})I_x(\bar{\nu})] d\bar{\nu}
\end{equation}
Here, the non-dimensional time is $\bar{t}=\Omega_0 t/2$ and the memory functions $G_1(\bar{t})$ and $G_2(\bar{t})$ are given by

\begin{equation}
G_1(\bar{t})=\frac{\bar{t}}{(1+\bar{t}^2)^2}, \;\; G_2(\bar{t})=\frac{1}{2} \frac{1-\bar{t}^2}{(1+\bar{t}^2)^2}
\end{equation}

\subsubsection{The slution at $\mathcal{O}(1)$ Richardson numbers}

We define the Laplas transform of $I_x, I_z, G_1$ and $G_2$ in the same way as was done for the three-dimensional case. Then by applying the Laplas transform to equations (23), (24), we obtain the characteristic equation for the eigenvalues corresponding to the poles of the transformed components $I_x(s)$ and $I_z(s)$

\begin{equation}
s^2+4 Ri G_2((s)s+4Ri^2[G_1^2(s)+G_2^2(s)]+2 Ri-8 RiG_1(s)+\frac{4Ri^2}{s}G_2(s)=0
\end{equation}
For values of $Re(s) >0$, it is convenient to express the transformed memory functions $G_1(s)$ and $G_2(s)$ in terms of cosine (ci) and sine (si) integrals:

\begin{equation*}
G_1(s)=\frac{1}{2} {1-s[ci(s)sin(s)-si(s)cos(s)]},
\end{equation*}
\begin{equation*}
G_2(s)=-\frac{1}{2} s[ci(s)cos(s)+si(s)sin(s)].
\end{equation*}
The corresponding values of $G_1(s)$ and $G_2(s)$ for $Re(s) \leq 0$ are obtained by using the procedure of analytical continuation.

The solution of equation (26) with $Re(s) \geq 0$ are shown in Figure 4 as function of the Richardson number.

\begin{figure}[h]
\centering
\includegraphics[scale=1.2]{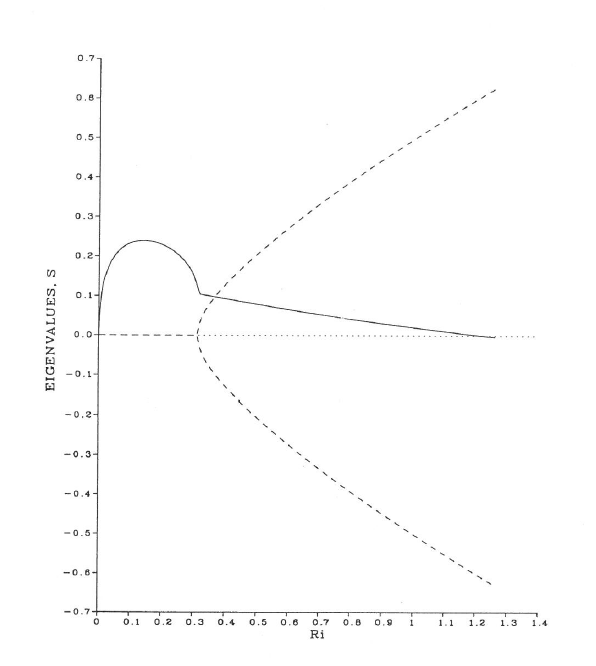}
\end{figure}

Figure 4. The dependence of the eigenvalues, which correspond to time growing disturbances, on the Richardson number for two-dimensional case. The real part of the eigenvalue ($Re(s)$) is shown by the solid curve while the dash line corresponds to its imaginary part ($Im(s)$). The dotted line corresponds to the marginal value for stability.
\\
\\
As can be seen, the homogeneous shear flow without stratification ($Ri=0$) does not display the exponential instability (although there is still the possibility of algebraic grows of the disturbance, as was shown by Landhal, 1970). However, contrary to intuition, the addition of stably stratification has a destabilizing effect.

Indeed, in the three-dimensional case the mechanism of localized vortex instability is associated with its interaction with the basic shear flow. The growth rate of fluid impulse of the disturbance has a maximum at $Ri=0$ and only decreases as the Richardson number increases. In contrast, in the two-dimensional case the buoyancy, which is solely responsible for new vorticity generation, gives rise to exponential instability up to $Ri \sim 1.23$. Nevertheless, for values of $Ri > 0.3$ the solution exhibits oscillatory behavior similar to the three-dimensional case. 

It should be noted that for the two-dimensional case, interpretation of the relationship between the fluid impulse and the vorticity field associated with the localized disturbance is more difficult than for three-dimensional case. As discussed above, the disturbance is considered to be a compact vortex with vorticity tails, the strength of which is rapidly reduced with distance from the disturbed region. Indeed, in two-dimensional case the numerical simulations by Robius $\And$ Delisi (1990) and Carten {\itshape et al.} (1998) demonstrated that the motion of the vortex pair in a stratified environment causes the generation of an extended wake of vorticity. Thus, the flow pattern in this case may not comply with the requirement for compactness of the disturbance. For this requirement to be satisfied, the self-induced displacement of the vortex pair during the typical evolution tme of the fluid impulse must be less than its typical dimension. This condition can be expressed as

\begin{equation*}
Fr=\frac{W}{Nb}<<1,
\end{equation*}
where $Fr$ is the non dimensional Froude number, b is the vortex core separation distance and $W$ is the magnitude of the vertical velocity induced by the vortex pair. In this respect, it is unfortunate that the present theoretical results can not be compared with the numerical results of  Robius $\And$ Delisi (1990) and Carten {\itshape et al.} (1998) as their simulations were carried out only for range of $Fr \geq 1$.

\subsubsection{The solution at large Richardson numbers}

In contrast to the three-dimensional case, the solution at $Ri>>1$ is physically meaningful, because the fact that the subdivision procedure used here has no need to add of the artificial vorticity field within the disturbed region. Taking limit $\Omega \to 0$ and turning to the dimensional time in (23) and (24), the evolution equations for fluid impulse components can be written as

\begin{equation}
\frac{dI_x}{dt}=-N^2 \int_0^{t} I_x(\tau) d\tau 
\end{equation}
\begin{equation}
\frac{dI_z}{dt}=-3 N^2 \int_0^{t} I_z(\tau) d\tau 
\end{equation}

Equations (27), (28) describe the independent oscillations in $x$ and $z$ directions. i.e.

\begin{equation}
I_x(t)=I_x(0) \cdot cos(Nt) \;\; and \;\; I_z(t)=I_z(0) \cdot cos(Nt/\sqrt{3})
\end{equation}
For comparison, Saffman's analysis (Safman, 1972) gives an oscillating motion in the $z$ direction with frequency of about $N/\sqrt{2.2}$. The main source of this discrepancy shows up also in comparison with analysis by Spalart (1996). 

Spalart (1996) suggested to subdivide the whole vorticity field into three parts. He introduced the notion $\omega_1$ for the initially created vorticity, $\omega_2$ for the vorticity field generated by buoyancy in the disturbed region and its surrounding and $\omega_3$ for the generated vorticity tails. Then he ignored the $\omega_3$ vorticity field and used the fluid impulse determined by expression (19). Evolution equations obtained by him for fluid impulse components can be written in our notations as 

\begin{equation}
\frac{dI_x}{dt}=0 \;\; and \;\; \frac{dI_x}{dt}=-\frac{g}{\rho_0} \int \rho dA,
\end{equation}
where $\rho$ is the disturbed density.

As was discussed above, the magnitude of the fluid impulse integral defined by expression (19) depends on way the volume of integration is allowed to tend to infinity. Or, in other words, it depends on how the generated vorticity tails are taken into account. Saffman's (1972) suggestion was to consider the motion of the vortex pair in a stratified atmosphere as a sequence of states, each being described by a motion of vortex through a uniform field. In this case the vorticity tails are not generated at all, but the correspondence of this approximation solution to the starting problem stands unproved. The Spalart's (1996) suggestion to ignore the vorticity tails can be accepted only on the basis of well-determined mathematical procedure. Moreover, according to (7), even though the disturbed vorticity is localized, density distribution possesses far field asymptotic tails $\mathcal{O}(1/|\mathbf{r}|^2)$. Hence the integral in the right hand side of (30) is not absolutely convergent and thus the problem remains unsolved.

We define the subdivision procedure in such a way as to extract the localized part of the disturbance without changing its fluid impulse. The latter requirement determines the corresponding mathematical procedure, which in case according to (22) resulted in the following set of evolution equations

\begin{equation}
\frac{dI_x}{dt}=-\frac{g}{\rho_0} \lim_{R \to \infty} \int_{|\mathbf{r}| \leq R} z \frac{\partial \rho^I}{\partial x} dA
\end{equation}
\begin{equation}
\frac{dI_z}{dt}=\frac{g}{\rho_0} \lim_{R \to \infty} \int_{|\mathbf{r}| \leq R} x \frac{\partial \rho^I}{\partial x} dA
\end{equation}
where the region of integration is a circle of infinite radius. Subsequent calculations reduce (31), (32) to equations (27), (28).

\section{Summary and discussion}

In the present paper the evolution of localized vortex disturbance in stably stratified shear flows was analyzed using the fluid impulse concept. A set of equations describing the temporal evolution of the components of the fluid impulse was obtained for flows associated with two- and three dimensional disturbances. The corresponding sets of equations (23), (24) and (14), (16) have an integro-differential character where the terms describing the effect of stratification appear as convolution integral of the fluid impulse and the time depended 'memory' functions. These functions ae calculated for the case where the external parallel shear flow varies only in the direction of the gravitational force. 

From the solution of evolution equations it follows that in both the two- and the three-dimensional cases there is a range of Richadson numbes where the fluid impulse associated with the disturbance grows exponentially. The respective upper limits of the range are $Ri \sim 1.23$ and $Ri \sim 0.89$. In addition, both cases are also characterized by a critical value of Richardson number (around $Ri \sim 0.3$ for both cases), beyond which the solution exhibits oscillatory behavior. This oscillatory behavior is a result of two opposite effects: the growth of the vortex due to its movement in the direction of the mean shear and the opposite gravitational force effect which is the result of the accumulation of the disturbed mass during the movement of the vortex. For lower values of $Ri$ the stretching of the vortex by the shear is strong enough to overcome the effect of gravitational force.

Finally, we correlate the predictions of the model with experimental and numerical results concerning the stably stratified turbulent shear flows. Two main criteria have been employed to distinguish between the shear and buoyancy dominated flows. The first one is the stationary Richardson number ($Ri_s$) at which turbulence growth is suppressed (Rohr {\itshape et al.}, 1988). A comparison of the present results with this criterion can not be made because there is no one-to-one correspondence between growth of the fluid impulse and growth of the energy of the localized vortex. The latter strongly depends on viscosity (it was found experimentally that $Ri_s$ grows with the Reynolds number), whereas the fluid impulse of the disturbance is not affected by viscosity (LC).

The second criterion was introduced by Holt {\itshape et al.} (1992). They defined the transition Richardson number as the value at which the vertical density flux (as well as other parameters of the flow) begin to exhibit time oscillations. This qualitative change in the character of the flow can be directly related to the time evolution of the fluid impulse associated with the localized disturbance. The comparison shows that the corresponding value of the Richardson number obtained in the present study, $Ri \sim 0.29$, is close to the transition Richardson number ($Ri = 0.25$) reported by Holt {\itshape et al.} (1992). It is important to emphasize that the critical Richardson number found here can explain the appearance of the oscillatory character of the flow without relating it to a 'linear instability' of wavy disturbances. In fact, the oscillatory behavior of the flow (over a certain range of the Richardson numbers) is an inherent feature of a localized vortex embedded in a stably stratified shear flow.

I thank professor Jacob Cohen for his interest and helpful comments and suggestions. The work was supported by Israel Ministry of Immigrant Absorbtion.

\begin{appendices}
\setcounter{equation}{0}
\renewcommand\theequation{A\arabic{equation}}
\section{}

This Appendix describes the subdivision procedure and the calculation of the derivative of (10). In order to extract the localized part of the vorticity field the whole space is subdivided into two regions, inside and outside a spherical domain of radius $R$, enclosing the disturbance. In the outer region $|\mathbf{x}| \geq R$, the corresponding vorticity and density are set to satisfy the following equations
\begin{equation*}
\frac{\partial \boldsymbol{\omega}^I}{\partial t}+ (\mathbf{U}_{0} \cdot \mathbf{\nabla}) \boldsymbol{\omega}^I-(\boldsymbol{\omega}^I \cdot \mathbf{\nabla}) \mathbf{U}_{0}-(\boldsymbol{\Omega}_0 \cdot \mathbf{\nabla})( \mathbf{u}^I-\mathbf{u}^a)+
\end{equation*}

\begin{equation}
 (\mathbf{u} \cdot \mathbf{\nabla}) \boldsymbol{\omega}-(\boldsymbol{\omega} \cdot \mathbf{\nabla}) \mathbf{u}=\frac{1}{\rho_0} \mathbf{\nabla} (\rho^I-\rho^a) \times \mathbf{g} 
\end{equation}

\begin{equation}
\frac{\partial \boldsymbol{\omega}^{II}}{\partial t}+ (\mathbf{U}_{0} \cdot \mathbf{\nabla}) \boldsymbol{\omega}^{II}-(\boldsymbol{\omega}^{II} \cdot \mathbf{\nabla}) \mathbf{U}_{0}-(\boldsymbol{\Omega}_0 \cdot \mathbf{\nabla}) (\mathbf{u}^{II}+\mathbf{u}^a)=\frac{1}{\rho_0} \mathbf{\nabla} (\rho^{II}+\rho^a) \times \mathbf{g} 
\end{equation}

\begin{equation}
\frac{\partial \rho^I}{\partial t}+ (\mathbf{U}_{0} \cdot \mathbf{\nabla}) \rho^I+ (\mathbf{u} \cdot \mathbf{\nabla}) \rho+u_z^I \frac{d\rho_0}{dz}=0, 
\end{equation}

\begin{equation}
\frac{\partial \rho^{II}}{\partial t}+ (\mathbf{U}_{0} \cdot \mathbf{\nabla}) \rho^{II}+u_z^{II} \frac{d\rho_0}{dz}=0, 
\end{equation}
Here $\mathbf{u}^a$ is the leading term of the far-field asymptotic expansion of the vorticity field induced by the $\boldsymbol{\omega}^I$, i.e.

\begin{equation}
\mathbf{u}^a(\mathbf{x},t)=\frac{1}{4\pi} [\frac{\mathbf{I(t)}}{|\mathbf{x}|^3}-\frac{3(\mathbf{I(t)} \cdot \mathbf{x})\mathbf{x}}{|\mathbf{x}|^5}]
\end{equation}
where $\mathbf{I(t)}$ is given by (10). Similarly, $\rho^a$ is the leading term of the far-field asymptotic expansion of the density distribution which can be obtained by solving (35)

\begin{equation}
\rho^{a}(\mathbf{x},t)=- \frac{d\rho_0}{dz} \int_0^t d\tau u_z^a(x-\Omega_0(t-\tau)z,y,z,\tau) 
\end{equation}

In the inner region $|\mathbf{x}| < R$, the corresponding vorticity and density satisfy the following equations:

\begin{equation*}
\frac{\partial \boldsymbol{\omega}^I}{\partial t}+ (\mathbf{U}_{0} \cdot \mathbf{\nabla}) \boldsymbol{\omega}^I-(\boldsymbol{\omega}^I \cdot \mathbf{\nabla}) \mathbf{U}_{0}-(\boldsymbol{\Omega}_0 \cdot \mathbf{\nabla}) \mathbf{u}^I+ (\mathbf{u} \cdot \mathbf{\nabla}) \boldsymbol{\omega}-
\end{equation*}

\begin{equation}
(\boldsymbol{\omega} \cdot \mathbf{\nabla}) \mathbf{u}=\frac{1}{\rho_0} \mathbf{\nabla} \rho^I \times \mathbf{g} -\mathbf{\nabla} \Psi
\end{equation}

\begin{equation}
\frac{\partial \boldsymbol{\omega}^{II}}{\partial t}+ (\mathbf{U}_{0} \cdot \mathbf{\nabla}) \boldsymbol{\omega}^{II}-(\boldsymbol{\omega}^{II} \cdot \mathbf{\nabla}) \mathbf{U}_{0}-(\boldsymbol{\Omega}_0 \cdot \mathbf{\nabla}) \mathbf{u}^{II}=\frac{1}{\rho_0} \mathbf{\nabla} \rho^{II} \times \mathbf{g} +\mathbf{\nabla} \Psi
\end{equation}

\begin{equation}
\frac{\partial \rho^I}{\partial t}+ (\mathbf{U}_{0} \cdot \mathbf{\nabla}) \rho^I+ (\mathbf{u} \cdot \mathbf{\nabla}) \rho+u_z^I \frac{d\rho_0}{dz}=0, 
\end{equation}

\begin{equation}
\frac{\partial \rho^{II}}{\partial t}+ (\mathbf{U}_{0} \cdot \mathbf{\nabla}) \rho^{II}+u_z^{II} \frac{d\rho_0}{dz}=0, 
\end{equation}
It should be emphasized that the sum of the equations describing the time evolution of $\boldsymbol{\omega}^I$ and $\boldsymbol{\omega}^{II}$ as well as $\rho^I$ and $\rho^{II}$ in each equation yields equation (6) and (7) for that region, and together with the initial conditions given in (9), yields the original problem for the entire space.

The divergence free condition (8) for the vorticity fields $\boldsymbol{\omega}^I$ and $\boldsymbol{\omega}^{II}$ is always satisfied in the outer region. For this condition to be satisfied in the inner region, $\mathbf{\nabla}^2 \Psi$ must be equal to zero, as can be shown by applying the operator $(\mathbf{\nabla} \cdot)$ to (A7) and (A8). Then $\Psi$ is determined by solving the Neumann problem for which the normal derivative of $\Psi$ at $|\mathbf{x}| = R$ is matched with the scalar product of the unit vector normal to the boundary surface $\mathbf{n}$, and the term in (A7) containing $\mathbf{u}^a$ and $\rho^a$, i.e.

\begin{equation}
\mathbf{n} \cdot \frac{\partial \Psi}{\partial\mathbf{n}}|_{|\mathbf{x}| = R}=\mathbf{n} \cdot [ (\mathbf{\Omega}_{0} \cdot \mathbf{\nabla})\mathbf{u}^a+\frac{1}{\rho_0} \mathbf{\nabla}\rho^a  \times \mathbf{g})|_{|\mathbf{x}| = R}
\end{equation}

The exact solution for $\Psi$ in the inner region is not relevant for the present work. To satisfy (8) it is necessary to know that such a solution exist. The latter follows from the fact that the expression within the square brackets can be written as a rotor of another vector function and, thus, both necessary and sufficient conditions for the existence of the solution for $\Psi$ are satisfied. 

Since the additional terms with $\rho^a$ and $\mathbf{u}^a$ in (A1) cancel the leading terms of the far-field vorticity generated via the problematic terms (see the discussion following equations 6-7), the asymptotic behavior of $\boldsymbol{\omega}^I$ is described by

\begin{equation}
|\boldsymbol{\omega}^I( \mathbf{x},t)| \sim  \mathcal{O}(1/|\mathbf{x}|^5) \;\; for \;\; |\mathbf{x}| >> l
\end{equation}
\setcounter{equation}{0}
\renewcommand\theequation{B\arabic{equation}}
\section{}
The purpose of this Apendix is to evaluate the last integral on the right hand side of equation (13), i.e.

\begin{equation}
 \mathbf{J}=\frac{1}{2\rho_0} \lim_{R_1 \to \infty} \int_{|\mathbf{x}| \leq R_1} \mathbf{x} \times [\mathbf{\nabla} \rho^I \times \mathbf{g}]dV 
\end{equation}
Using integration by parts and employing Gauss' divergence theorem, the components of $\mathbf{J}$ in Cartesian tensor notations can be written as:

\begin{equation}
J_x=-\frac{g}{2\rho_0} \lim_{R_1 \to \infty}[ R_1 \oint_{|\mathbf{x}| = R_1} n_x n_z \rho^I dS]
\end{equation}

\begin{equation}
J_y=-\frac{g}{2\rho_0} \lim_{R_1 \to \infty}[ R_1 \oint_{|\mathbf{x}| = R_1} n_y n_z \rho^I dS]
\end{equation}

\begin{equation}
J_z=\frac{g}{2\rho_0} \lim_{R_1 \to \infty}[ R_1 \oint_{|\mathbf{x}| = R_1} (1- n_z^2)\rho^I dS]-2\int_{|\mathbf{x}| \leq R_1} \rho^I dV
\end{equation}
where $\mathbf{n}$ is a unit vector normal to the sphere surface. The last volumetric integral in (B4) is transformed into a surface by taking its time derivative and substituting the expression for $\partial \rho^I/\partial t$ from equation (A3). This yields

\begin{equation}
\int_{|\mathbf{x}| \leq R_1} \rho^I dV=-\int_0^t d\tau \oint_{|\mathbf{x}| = R_1} n_xn_zR_1\Omega_0\rho^IdS-\frac{2}{3}\frac{d\rho_0}{dz}\int_0^tI_z(\tau)d\tau
\end{equation}

Each one of the surface integrals in (B2)-(B5) can be evaluated for a finite value of $R_1$ and then its limit as $R_1 \to \infty$ can be obtained. In these integrals, the leading term of the far-field expression for $\rho^I$ is obtained by substituting (A5) into (A6).

\begin{equation*}
\rho^{I}(\mathbf{x},t)|_{|\mathbf{x}| \to \infty}=\frac{1}{4\pi} \frac{d\rho_0}{dz} \int_0^t d\tau [ \frac{I_z(\tau)}{[(x-\Omega_0 (t-\tau)z)^2+y^2+z^2]^{3/2}}-
\end{equation*}

\begin{equation}
 \frac{3 z (I_x(\tau)(x-\Omega_0 (t-\tau)z)+I_y(\tau) y+I_z(\tau) z}{[(x-\Omega_0 (t-\tau)z)^2+y^2+z^2]^{5/2}})]
\end{equation}
Since $\rho^I$ decays like $1/|\mathbf{x}|^3$, it gives a non-vanishing contribution to the integrals (B2)-(B5), i.e.

\begin{equation}
J_x=N^2 \int_0^t [G_{xx}(t-\tau)I_x(\tau)+G_{xz}(t-\tau)I_z(\tau)]d\tau
\end{equation}

\begin{equation}
J_y=N^2 \int_0^t G_{yy}(t-\tau)I_y(\tau) d\tau
\end{equation}

\begin{equation*}
J_z=N^2 \int_0^t [G_{zx}(t-\tau)I_x(\tau)+G_{zz}(t-\tau)I_z(\tau)]d\tau-\frac{2}{3}\int_0^t I_z(\tau) d\tau-
\end{equation*}
\begin{equation}
2N^2 \Omega_0 \int_0^t d\tau \int_0^{\tau} dv[G_{xx}(t-\tau)I_x(\tau)+G_{xz}(t-\tau)I_z(\tau)]
\end{equation}
Here $N$ is the Brunt-V$\ddot{a}$is$\ddot{a}$l$\ddot{a}$ frequency defined as 
\begin{equation}
N^2=-\frac{g}{\rho_0}\frac{d\rho_0}{dz}
\end{equation}
and $G_{xx}, G_{yy}, G_{xz}, G_{zx}$ and $G_{zz}$ are the memory functions, the dependence of which on the non-dimensional time $\bar{t}=\Omega_0 t$ is shown in Figure 2.

\end{appendices}

\end{document}